
\def\xmz{\hat x_\mu^{(0)}}\def\pmz{\hat p_\mu^{(0)}}\def\tG{{\tilde G}}
\def\cO{{\cal O}}\def\cL{{\cal L}}
\def\ref#1{\medskip\everypar={\hangindent 2\parindent}#1}
\def\beginref{\begingroup
\bigskip
\centerline{\bf References}
\nobreak\noindent}
\def\endref{\par\endgroup}

\def\PL#1{Phys.\ Lett.\ {\bf#1}}

\def\PR#1{Phys.\ Rev.\ {\bf#1}}\def\CQG#1{Class.\ Quantum Grav.\ {\bf#1}}

\def\JMP#1{J.\ Math.\ Phys.\ {\bf#1}}

\def\JHEP#1{JHEP\ {\bf#1}}
\def\RMP#1{Rev.\ Mod.\ Phys.\ {\bf#1}}

\def\arx#1{{\tt arXiv:#1}}

\def\hx{{\hat x}}\def\hp{{\hat p}}\def\xp{\,x{\cdot}p\,} \def\px{\,p{\cdot}x\,}
\def\lra{\leftrightarrow}
\def\ha{{1\over2}}\def\diag{{\rm diag}}
\def\section#1{\bigskip\noindent{\bf#1}\smallskip}

\def\pb{Poisson brackets }\def\bdot{{\cdot}}\def\inf{\infty}\def\sn{\mathop{\rm sn}\nolimits}

{\nopagenumbers
\line{}
\vskip40pt
\centerline{\bf Realizations of the Yang-Poisson model on canonical phase space}
\vskip50pt
\centerline{{\bf S. Meljanac}\footnote{$^\dagger$}{e-mail: meljanac@irb.hr}}
\vskip5pt
\centerline {Rudjer Bo\v skovi\'c Institute, Theoretical Physics Division}
\centerline{Bljeni\v cka c.~54, 10002 Zagreb, Croatia}
\vskip10pt
\centerline{{\bf S. Mignemi}\footnote{$^\ast$}{e-mail: smignemi@unica.it}}
\vskip5pt
\centerline {Dipartimento di Matematica, Universit\`a di Cagliari}
\centerline{via Ospedale 72, 09124 Cagliari, Italy}
\smallskip
\centerline{and INFN, Sezione di Cagliari}
\centerline{Cittadella Universitaria, 09042 Monserrato, Italy}

\vskip40pt
\centerline{\bf Abstract}
\medskip
{\noindent We discuss exact realizations of the Yang-Poisson model on canonical phase space.
The Yang model is an example of noncommutative geometry on a background spacetime of constant curvature
and is notable for its duality between position and momentum manifolds.
We call Yang-Poisson model its classical limit, with commutators replaced by Poisson brackets.
The structure is simpler in the classical case, and exact realizations can be found.
}
\vskip60pt
%P.A.C.S. Numbers: 04.60.-m 04.70.Bw 11.15.-q
\vfil\eject}

\section{1. The Yang-Poisson model}
The Yang model of noncommutative geometry on a curved background spacetime [1] was proposed soon after the invention
of noncommutative geometry by Snyder [2], and is a generalization of that model.
It is defined in terms of a 15-parameter algebra isomorphic to $so(1,5)$, that contains the generators of the Lorentz
algebra together with the coordinates of phase spaces.
Such model was slightly generalized in [3].

Its interest lies in the preservation of Lorentz invariance in spite of the granular structure of
spacetime due to noncommutativity and in the assumption of a constant curvature background manifold, entailing a
duality between the curvature of position and momentum space.
It also gave rise to a model where the same symmetry is realized nonlinearly, called Snyder-de Sitter or triply special
relativity, that has received some attention in the literature [4].

Quantum realizations of the Yang model on different Hilbert spaces have been discussed in several papers in recent
times [5-9].
The results of these investigation seem to imply that it is not possible to obtain realizations in closed analytic
form, but only as power series in the coupling constants.

However, the classical limit of the algebra, obtained replacing the quantum commutators by Poisson brackets, has a
simpler structure
and is much easier to solve since no ordering problems arise, so that the realizations can be obtained in an exact
form, as we shall show in this paper.
In the following, for obvious reasons, we shall refer to this  model as Yang-Poisson model.

The Yang-Poisson model is generated by a Poisson algebra containing the usual Lorentz algebra of generators $M_{\mu\nu}$,
with its standard action on a phase space parametrized by $\hx_\mu$ and $\hp_\mu$,\footnote{$^1$}{We use the conventions
$\mu,\nu=0,\dots,n-1$, $\eta_{\mu\nu}=\diag(-1,1,\dots,1)$.}
$$\{M_{\mu\nu},M_{\rho\sigma}\}=\big(\eta_{\mu\rho}M_{\nu\sigma}-\eta_{\mu\sigma}M_{\nu\rho}-\eta_{\nu\rho}M_{\mu\sigma}+\eta_{\nu\sigma}M_{\mu\rho}\big),\eqno(1)$$
$$\{M_{\mu\nu},\hx_\l\}=(\eta_{\mu\l}\hx_\nu-\eta_{\nu\l}\hx_\mu),\qquad\{M_{\mu\nu},\hp_\l\}=(\eta_{\mu\l}\hp_\nu-\eta_{\nu\l}\hp_\mu),\eqno(2)$$
together with a deformation of the Heisenberg algebra,
$$\{\hx_\mu,\hx_\nu\}=\beta^2M_{\mu\nu},\qquad\{\hp_\mu,\hp_\nu\}=\alpha^2M_{\mu\nu},\eqno(3)$$
$$\{\hx_\mu,\hp_\nu\}=\eta_{\mu\nu} h,\eqno(4)$$
and a further scalar generator $h$, necessary to close the algebra, satisfying
$$\{h,\hx_\mu\}=\beta^2\hp_\mu,\qquad\{h,\hp_\mu\}=-\alpha^2\hx_\mu,\eqno(5)$$
$$\qquad\{M_{\mu\nu},h\}=0.\eqno(6)$$
Here $\alpha$ and $\beta$ are real parameters and $\eta_{\mu\nu}$ the flat metric.
Well-known limit cases are obtained for $\alpha\to0$ (Snyder algebra) and $\beta\to0$ (de Sitter algebra). The \pb  satisfy
the Jacobi identities, and a Born duality [9] holds for $\alpha\lra\beta$, $\hx_\mu\to-\hp_\mu$,  $\hp_\mu\to\hx_\mu$,
$M_{\mu\nu}\lra M_{\mu\nu}$, $h\lra h$.

In the following  we shall discuss the realizations of the Yang-Poisson algebra on a canonical phase space of coordinates
$x_\mu$ and $p_\mu$, with the Lorentz algebra generators realized as $M_{\mu\nu}=x_\mu p_\nu-x_\nu p_\mu$.
We shall obtain realizations of the algebra by solving the differential equations related to the Poisson brackets,
and then will rederive them by explicitly summing a perturbative expansion in $\alpha^2$ and $\beta^2$.
To illustrate a simple application of the formalism,
we also briefly consider the modification of the classical solutions of the one-dimensional harmonic oscillator.
These results can be useful in the study of the quantum problem.

\section{2. Special solution}
As mentioned above, we look for realizations of $\hx_\mu$ and $\hp_\mu$ on a phase space with coordinates $x_\mu$ and $p_\mu$
satisfying the canonical algebra
$$\{x_\mu,x_\nu\}=\{p_\mu,p_\nu\}=0,\qquad\{x_\mu,p_\nu\}=\eta_{\mu\nu}.\eqno(7)$$

We make the ansatz, suggested by previous investigations of the Yang model [5-7],
$$\hx_\mu=f(p^2,z)\,x_\mu ,\qquad \hp_\mu=g(x^2,z)\,p_\mu ,\eqno(8)$$
$$M_{\mu\nu}=x_\mu p_\nu-x_\nu p_\mu,\qquad h=h(x^2,p^2,z),\eqno(9)$$
where $z=\alpha\beta\, x\bdot p$ and $f$ and $g$ are functions to be determined.

The only nontrivial brackets to be checked are (3)-(4). These give rise to partial differential equations.
From eqs.~(3) it follows that
$$f=\sqrt{1-\beta^2p^2+\phi_1(z)},\qquad g=\sqrt{1-\alpha^2x^2+\phi_2(z)}.\eqno(10)$$
with arbitrary functions $\phi_1$ an $\phi_2$, while, from eq.~(4),
$$\phi_1\phi_2+\phi_1+\phi_2=z^2,\qquad h=fg.\eqno(11)$$
In particular, we can assume that the realization is symmetric under the exchange of $x$ and $p$, as is natural in
view of the Born duality of the model. In this case, $\phi_1=\phi_2=\phi$, and we obtain
$$\phi=\sqrt{1+z^2}-1,\eqno(12)$$
and then
$$f=\sqrt{\sqrt{1+z^2}-\beta^2p^2},\qquad g=\sqrt{\sqrt{1+z^2}-\alpha^2x^2},\eqno(13)$$
This gives an exact realization of the Yang model, symmetric for $x\lra p$ and $\alpha\lra\beta$.
The same result can be obtained (less explicitly) by an expansion in powers of $\alpha^2$ and $\beta^2$, as reported in appendix B.

More generally, we can set
$$\phi_1(z)=c_1\psi(z),\qquad\phi_2(z)=c_2\psi(z)\eqno(14)$$
with $c_1+c_2=1$. Then we get
$$c_1c_2\psi^2+\psi-z^2=0,\eqno(15)$$
with solution
$$\psi(z)={\sqrt{1+4c_1c_2z^2}-1\over2c_1c_2},\eqno(16)$$
and then
$$\eqalignno{&\phi_1(z)={\sqrt{1+4c_1(1-c_1)z^2}-1\over2(1-c_1)}={2c_1z^2\over\sqrt{1+4c_1(1-c_1)z^2}+1},\cr
&\phi_2(z)={\sqrt{1+4c_2(1-c_2)z^2}-1\over2(1-c_2)}={2c_2z^2\over\sqrt{1+4c_2(1-c_2)z^2}+1}.&(17)}$$
It follows
$$\hx_\mu(c_1)=\sqrt{1-\beta^2p^2+\phi_1(z)}\ x_\mu,\qquad \hp_\mu(c_2)=\sqrt{1-\alpha^2x^2+\phi_2(z)}\ p_\mu,\eqno(18)$$
and
$$h(c_1,c_2)=\sqrt{\left[1-\beta^2p^2+\phi_1(z)\right]\left[1-\alpha^2x^2+\phi_2(z)\right]}.\eqno(19)$$
It is easy to check that (5) is satisfied. Moreover, one can show that in terms of the original variables,
$$h=\sqrt{1-\alpha^2\hx^2-\beta^2\hp^2-{\alpha^2\beta^2\over2}M^2}.\eqno(20)$$
This result for $h$ is universal and generally valid in the Yang-Poisson model.
This follows from the application of the group of automorphisms defined in section 3.
\bigbreak

\section{3. General solution}
We now use an alternative method to calculate the realizations of $\hx_\mu$ and $\hp_\mu$, analogous to the one used in ref.~[9]
for the quantum case.
We define
$$\xmz=\sqrt{1-\beta^2p^2}\ x_\mu,\qquad\pmz=\sqrt{1-\alpha^2x^2}\ p_\mu,\eqno(21)$$
and then construct an operator $\cO$ such that
$$\{\cO(\xmz),\hat p_\nu^{(0)}\}=\eta_{\mu\nu} h,\eqno(22)$$
so that in the right hand side appear only terms proportional to $\eta_{\mu\nu}$.

The general structure of the operator $\cO$ acting on $\xmz$ is
$$\cO(\xmz)=(\exp\cL_G)(\xmz)=\xmz+\{G,\xmz\}+{1\over2!}\{G,\{G,\xmz\}\}+{1\over3!}\{G,\{G,\{G,\xmz\}\}\}+\dots,\eqno(23)$$
where the operator $\cL_G$ is defined as
$$\cL_G\big(F(x,p)\big)=\{G,F(x,p)\},\eqno(24)$$
for any function $F(x,p)$.
Solving perturbatively (22) we find a unique solution for $G$,
$$G=\sum_{n=1}^\inf\alpha^{2n}\beta^{2n}g_{2n},\eqno(25)$$
with
$$g_{2n}=Q_n{(\xp)^{2n+1}\over2n+1},\qquad Q_n={(-1)^n\over2n}.\eqno(26)$$
The sum of the series (25) gives an exact result,
$$G={1\over\alpha\beta}\left[z\left(1-\ha\ln(1+z^2)\right)-\arctan z\right].\eqno(27)$$

We now perform the transformation (23) with the function $c_1G$, where $c_1$ is a real constant, obtaining
$$\eqalignno{\hx_\mu(c_1)&=\big(\exp(c_1\cL_G)\big)(\xmz)\cr&=\xmz+c_1\{G,\xmz\}+{c_1^2\over2!}\{G,\{G,\xmz\}\}+{c_1^3\over3!}\{G,\{G,\{G,\xmz\}\}\}+\dots&(28)} $$
For $c_1=0$ we have
$$\hx_\mu(0)=\xmz=\sqrt{1-\beta^2p^2}\,x_\mu.\eqno(29)$$

Substituting (25)-(26) into (28), we get after some manipulations
$$\eqalignno{\hx_\mu(c_1)&=\sum_{m=0}^\inf\sum_{k=0}^\inf{\ha\choose m}\bigg[c_1(2m-1)Q_k+c_1^2\ {(2m-1)^2\over2!}\sum_{k_1+k_2=k}Q_{k_1}Q_{k_2}\cr
&+c_1^3\ {(2m-1)^3\over3!}\sum_{k_1+k_2+k_3=k}Q_{k_1}Q_{k_2}Q_{k_3}+\dots+c_1^k\ {(2m-1)^k\over k!}\ Q_1^k\bigg]\ z^{2k}(-\beta^2p^2)^mx_\mu\cr
&=\sum_{m=0}^\inf\sum_{k=0}^\inf{\ha\choose m}{{(1-2m)c_1\over 2}\choose k}(-\beta^2p^2)^mz^{2k}x_\mu.&(30)}$$
The sum of this series gives again (18) with $\phi_1$ given by (17).

\bigskip
Let us now define
$$\eqalignno{\hp_\mu(c_2)&=\big(\exp(-c_2\cL_G)\big)(\pmz)\cr&=\pmz-c_2\{G,\pmz\}+{(-c_2)^2\over2!}\{G,\{G,\pmz\}\}+{(-c_2)^3\over3!}\{G,\{G,\{G,\pmz\}\}\}+\dots&(31)}$$
For $c_2=0$ we have
$$\hp_\mu(0)=\pmz=\sqrt{1-\alpha^2x^2}\,p_\mu.\eqno(32)$$

Using $G$ defined above, we get
$$\eqalignno{\hp_\mu(c_2)&=\sum_{m=0}^\inf\sum_{k=0}^\inf{\ha\choose m}\bigg[c_2(2m-1)Q_k+c_2^2\ {(2m-1)^2\over2!}\sum_{k_1+k_2=k}Q_{k_1}Q_{k_2}\cr
&+c_2^3\ {(2m-1)^3\over3!}\sum_{k_1+k_2+k_3=k}Q_{k_1}Q_{k_2}Q_{k_3}+\dots+c_2^k\ {(2m-1)^k\over k!}\ Q_1^k\bigg]\ z^{2k}(-\alpha^2x^2)^mp_\mu\cr
&=\sum_{m=0}^\inf\sum_{k=0}^\inf{\ha\choose m}{{(1-2m)c_2\over 2}\choose k}(-\alpha^2x^2)^mz^{2k}p_\mu.&(33)}$$
Again, this sum gives exactly eq.~(18) with $\phi_2$ in (17).
\bigskip
If we now consider
$$\hx_\mu(c_1)=f(c_1)x_\mu ,\qquad\hp_\mu(c_2)=g(c_2)p_\mu,\eqno(34)$$
then (4) is satisfied if $c_1+c_2=1$, with $h=h(c_1,c_2)=f(c_1)g(c_2)$, and
we obtain the same results for $\hx_\mu(c_1)$ and $\hp_\mu(c_2)$ as in section 2.

Note that if we assume $\phi_1=c_1\psi$ and $\phi_2=c_2\psi$, we reobtain eq.~(15) and then the solution (16).
In particular, for $c_1=c_2=\ha$, we recover the solution (13).

The most general realizations of $\hx_\mu$ and $\hp_\mu$ are obtained using the group of automorphisms
applied to the special solution defined by $c_1=1$, $c_2=0$,
$$\hx_\mu(1)=\sqrt{1-\beta^2p^2+z^2}\,x_\mu,\qquad\hp_\mu(0)=\sqrt{1-\alpha^2x^2}\,p_\mu,\eqno(35)$$
and
$$h(1,0)=\sqrt{(1-\beta^2p^2+z^2)(1-\alpha^2x^2)},\eqno(36)$$
namely,
$$\hx_\mu=\cO_\tG(\hx_\mu(1)),\quad\hp_\mu=\cO_\tG(\hp_\mu(0)),\quad h=\cO_\tG(h(1,0)),\eqno(37)$$
where $O_\tG=\exp(\cL_\tG)$ and $\tG$ is an arbitrary function of $\alpha^2x^2$, $\beta^2p^2$ and $z$.

\section{4. Application to the harmonic oscillator}

It may be interesting to study the dynamics of the Yang-Poisson model.
We consider the simplest instance, given by a one-dimensional harmonic oscillator.
Although in one dimension the $\hx$--$\hx$ and $\hp$--$\hp$ brackets are trivial, nevertheless the solutions present the main
characteristic of the higher-dimensional isotropic oscillator, since the radial equation in spherical coordinates is similar to the
one-dimensional equation.

We consider the usual Hamiltonian for a harmonic oscillator with frequency $\omega$,
$$H=\ha (\hp^2+\omega^2\hx^2).\eqno(38)$$
The field equations following from (38) with the symplectic structure (4) are
$$\dot\hx=\{x,H\}=h\,\hp,\qquad \dot\hp=\{p,H\} =-h\,\omega^2\hx.\eqno(39)$$
In one dimension, $h$ in eq.~(20) reduces to $h=\sqrt{1-\alpha^2\hx^2-\beta^2\hp^2}$.
From eqs.~(39) follows ${\dot{\hx}\over \hp}=-\omega^2{\dot \hp\over \hx}$, and hence the conservation of the Hamiltonian,
$$\hp^2+\omega^2\hx^2=2E,\eqno(40)$$
with $E$ the total energy.
Using (40), the first equation (39)  becomes
$$\dot\hx=\sqrt{[1-\alpha^2\hx^2-\beta^2(2E-\omega^2\hx^2)](2E-\omega^2\hx^2)},\eqno(41)$$
and then
$$t-t_0=\int{d\hx\over\sqrt{[1-2E\beta^2-(\alpha^2-\beta^2\omega^2)\hx^2](2E-\omega^2\hx^2)}},\eqno(42)$$
The solution can be written in terms of elliptic integrals: for example, if $1-2\beta^2E>0$ and $\alpha^2-\beta^2\omega^2>0$,
setting $t_0=0$,
$$\hx={\sqrt{2E}\over\omega}\sn\left(\sqrt{1-2\beta^2E}\omega t,m\right),\qquad{\rm with}\qquad m={2E(\alpha^2-\beta^2\omega^2)\over\omega^2(1-2\beta^2E)}.\eqno(43)$$
The solutions are periodic like for the classical oscillator, but now the period $T$ depends on the energy $E$,
$$T={2\pi\over\omega}{K(m)\over\sqrt{1-2\beta^2E}}\sim{2\pi\over\omega}\left[1+\ha\left(\beta^2+{\alpha^2\over\omega^2}\right)E+\dots\right],\eqno(44)$$
while the motion is allowed for $|\hx|>{\sqrt{2E}\over\omega}$ as in classical mechanics.

\section{5. Discussion}
In this paper we have obtained exact realizations of the Yang-Poisson algebra on a
canonical phase space. The results are simpler than in the quantum case [9]: for example, contrary to that case,
$G$ in (27) is a function only of $z$. Our result can be considered as a limit of the quantum formalism for $\hbar\to0$.

In fact, the operator $G$ for the Yang algebra was defined as $[e^{iG}\xmz e^{-iG},\hat p_\nu^{(0)}]=i\eta_{\mu\nu} h$ [9],
and is closely related to the operator $G$ for the Yang-Poisson algebra introduced in (23).
The Yang algebra operator $G$  is expressed as a series expansion in $\alpha^2x^2$, $\beta^2p^2$ and $\alpha\beta D={\alpha\beta\over2}(\xp+\px)$
with $[x_\mu,p_\nu]=i\eta_{\mu\nu}$ [9].

 In the classical limit, $[x_\mu,p_\nu]=0$ and $\alpha\beta D$ goes to $z$. Then, in the series
expansion of $G$ all terms of type $(\alpha\beta)^m D^n$ with $m>n$ can be omitted. In this way one can obtain the operator $G$
for the Yang-Poisson algebra (25)-(26). In the same way we obtain realizations of $\hx_\mu(c_1)$, $\hp_\mu(c_2)$ and
$h(c_1,c_2)$, with $c_1+c_2=1$ for the Yang-Poisson algebra.

A simple application of the formalism is the discussion of the harmonic oscillator. It turns out that the motion is
deformed with respect to the classical case, but still  periodic.
This result could be useful in the study of the quantum harmonic oscillator.
It would also be interesting to extend the investigation of the dynamics to more complex systems, like the three-dimensional oscillator
or the hydrogen atom, and compare the results with those obtained in the related triply special relativity framework [4,11].

\section{Acknowledgments}
S.~Me. wishes to thank D. Svrtan for comments related to the identities in appendix. S.~Mi. acknowledges support from GNFM.

\beginref
\ref [1] C.N. Yang, \PR{72}, 874 (1947).
\ref [2] H.S. Snyder, \PR{71}, 38 (1947).
\ref [3] V.V. Khruschev and A.N. Leznov, Grav. Cosmol. {\bf 9}, 159 (2003).
\ref [4] J. Kowalski-Glikman and L. Smolin, \PR{D70}, 065020 (2004).
H.G. Guo, C.G. Huang and H.T. Wu, \PL{B663}, 270 (2008).
S. Mignemi, \CQG{26}, 245020 (2009).
R. Banerjee, K. Kumar and D. Roychowdhury, \JHEP{1103}, 060 (2011).
\ref [5] S. Meljanac and R. \v Strajn, SIGMA {\bf18}, 022 (2022).
\ref [6] S. Meljanac and S. Mignemi, \PL{B833}, 137289 (2022).
\ref [7] S. Meljanac and S. Mignemi, \JMP{64}, 023505 (2023).
\ref [8] J. Lukierski, S. Meljanac, S. Mignemi and A. Pachol, \arx{2212.02316}.
\ref [9] T. Martini\'c-Bila\'c, S. Meljanac and S. Mignemi, \arx{2305.04013}.
\ref [10] M. Born, \RMP{21}, 463 (1949).
\ref [11] S. Mignemi, \CQG{29}, 215019 (2012).
\endref

\bigbreak

\section{Appendix A}
\noindent In this appendix we prove the identity used in the last step in eqs.~(30) and (33).

For fixed positive integers $k$, $l\le k$, $k_i$ with $i=1,\dots l$ satisfying the conditions
$$\sum_{i=1}^l k_i=k,\eqno(A.1)$$
and
$$q_1=k_1,\quad q_2=k_1+k_2,\quad\dots\quad q_{l-1}=\sum_{i=1}^{l-1} k_i\le k-1,\eqno(A.2)$$
the following identities hold:
$$C_l=\sum{1\over k_1k_2\cdots k_l}={l\,!\over k}\sum{1\over q_1q_2\cdots q_{l-1}}=(-1)^{k-l}\left({d\over dx}\right)^l{x\choose k}\bigg|_{x=0},\eqno(A.3)$$
with summations satisfying the above conditions.
\bigskip
\noindent In fact,

for $l=1$,
$$C_1={1\over k},$$

for $l=2$,
$$C_2={1\over k}\sum\left({1\over k_1}+{1\over k_2}\right)={2\over k}\sum_{i=1}^{k-1}{1\over q_i},$$

and in general, for $l\le k$,
$$C_l={l\over k}\sum {1\over k_1k_2\cdots k_{l-1}}={l(l-1)\over k}\sum{1\over q_{l-1}k_1k_2\cdots k_{l-2}}=\dots
={l\,!\over k}\sum{1\over q_{l-1}q_{l-2}\dots q_1}.\eqno(A.4)$$
This proves the first equality in (A.3)

Moreover,
$$\left({d\over dx}\right)^l{x\choose k}\bigg|_{x=0}=\ {l\,!\over k}\sum{1\over q_1q_2\cdots q_{l-1}}\ =
\sum_{k_1+k_2+\dots+k_l=k} {1\over k_1k_2\cdots k_l},\eqno(A.5)$$
where the summation is performed with the condition $q_1 < q_2 < q_3.......< q_{l-1} \le k-1$,
which proves the second equality.
\bigbreak

\section{Appendix B}
In this section, we write down an alternative derivation of the result of the previous sections by an expansion in $\alpha$ and $\beta$
in the simplest case of a realization symmetric for the exchange of $x$ and $p$. This will turn out to be the power
expansion of the exact realization (13).

The results obtained in previous investigations, suggest to write the expansion as
$$\hx_\mu=x_\mu\sum_{m=0}^\inf\,\sum_{n=0}^\inf\alpha^{2m+n}\beta^nf_{m,n}\,p^{2m}z^n,
\qquad\hp_\mu=p_\mu\sum_{m=0}^\inf\,\sum_{n=0}^\inf\alpha^n\beta^{2m+n} f_{m,n}\,x^{2m}z^n,\eqno(B.1)$$
and
$$h=\sum_{m,m'=0}^\inf\,\sum_{n=0}^\inf\alpha^{2m+n}\beta^{2m'+n} h^n_{m,m'}\,x^{2m}p^{2m'}z^n,\eqno(B.2)$$
with $f_{00}=1$. Imposing (3), (4),
%$$\{\hx_\m,\hx_\n\}=\beta^2M_{\mu\nu},\qquad\{\hp_\m,\hp_\n\}=\a^2M_{\mu\nu},$$
with $M_{\mu\nu}=x_\mu p_\nu-x_\nu p_\mu$, one obtains at order $\alpha^{2m+n}\beta^n$ (or equivalently $\alpha^n\beta^{2m'+n}$)
$$\sum_{l=0}^{m+1}\,\sum_{k=0}^{n} l f_{ l,k}\,f_{m+1-l,n-k}=0,\eqno(B.3)$$
and from (5) at order $\alpha^{2m+n}\beta^{2m'+n}$
$$\eqalignno{\sum_{k=0}^{n+1}&\ (n+1)f_{m,k}\,f_{m',n-k+1}-2mkf_{m,n-k+1}\,f_{m',k}-2m'kf_{m,k}\,f_{m',n-k+1}\cr
&-4(m+1)(m'+1)\sum_{k=0}^{n-1}f_{m+1,k}\,f_{m'+1,n-k-1}=0,&(B.4)}$$
from which one can calculate the $f_{m,n}$.
Then, from (5) follows also
$$h^n_{m,m'}=\sum f_{m,k}\,f_{m',n-k}.\eqno(B.5)$$
It is easy to check that these results are the expansion of the solution (13) in powers of $\alpha^2$ and $\beta^2$.
Up to fourth order in $\alpha$ and $\beta$, they are identical to those obtained in the quantum case in [6].

\end

The first terms of the expansion can be written up to eighth order as
$$\hx_\mu=x_\mu\Big[-{\beta^2\over2}p^2-{\beta^4\over8}p^4+{\alpha^2\beta^2\over4}z^2-{\beta^6\over16}p^6+{\alpha^2\beta^4\over8}z^2p^2
-{5\beta^8\over128}p^8+{3\alpha^2\beta^6\over32}z^2p^4-{3\alpha^4\beta^4\over32}z^4\Big].\eqno(A.6)$$
while $\hp_\mu$ is then given by the same expression with $x\lra p$, $\alpha\lra\beta$.

The expression of $h$ is instead, at the same order,
$$\eqalignno{h&=-\ha(\alpha^2x^2+\beta^2p^2)-{1\over8}(\alpha^4x^4-2\alpha^2\beta^2x^2p^2+\beta^4p^4)+{\alpha^2\beta^2\over2}z^2-{1\over16}(\alpha^6x^6+\beta^6p^6)\cr
&+{\alpha^2\beta^2\over16}(\alpha^2x^4p^2+\beta^2x^2p^4)-{5\over128}(\alpha^8x^8+\beta^8p^8)+{\alpha^2\beta^2\over32}(\alpha^4x^6p^2+\beta^4x^2p^6)+{\alpha^4\beta^4\over64}x^4p^4\cr
&+{\alpha^2\beta^2\over16}z^2(\alpha^4x^4-2\alpha^2\beta^2x^2p^2+\beta^4p^4)-{\alpha^4\beta^4\over8}z^4.&(A.7)}$$

It is easy to check that these results are the expansion of the solution (13) in powers of $\alpha^2$ and $\beta^2$.
Up to fourth order in $\alpha$ and $\b$, they are identical to those obtained in the quantum case in [6].

\end